\documentstyle[11pt]{article}
\begin{document}
\begin{titlepage}
\vspace{-10mm}
\begin{flushright}
             February 2001
\end{flushright}
\vspace{12pt}
\begin{center}
\begin{large}             
{\bf Scalar-Scalar Bound State in Non-commutative Space}\\
\end{large}
\vspace{5mm}
{\bf M. Haghighat
\footnote{e-mail: mansour@cc.iut.ac.ir}}
{\bf ,F. Loran 
\footnote{e-mail: farhang@theory.ipm.ac.ir}}\\
\vspace{12pt} 
{\it Department of  Physics, Isfahan University of Technology \\ 
Isfahan,  IRAN, \\ 
Institute for Studies in Theoretical Physics and Mathematics\\ 
P. O. Box: 5746, Tehran, 19395, IRAN.} 
\vspace{0.3cm}
\end{center}
\abstract{Bethe-Salpeter equation in the non-commutative space for a scalar-scalar 
bound state is considered. It is shown that in the non-relativistic limit, 
the effect of spatial non-commutativity appears as if there exist a magnetic 
dipole moment coupled to each particle. 
\ 
\vfill}
\end{titlepage}

Non-commutativity of space-time has been recently a subject of intense interest both 
in quantum mechanics and quantum field theory \cite{witten}-\cite{chai1}. 
In this paper, we would like to study the effects of such a non-commutativity on 
the spectra of the bound state of two scalar particles. 
Bethe-Salpeter (BS) equation \cite{sal,nak} is the usual tool for computing, 
for instance, the electromagnetic form factors and relativistic spectra of two body 
bound states. In the following analysis we examine scalar-scalar bound state spectra.
\par
The BS equation for two scalar particles is 
\begin{eqnarray}\label{a1}
\Gamma(p_1,p_2)=\int d^4k I(k;p_1,p_2,\theta) D(p_1+k,p_2-k) \Gamma(p_1+k,p_2-k).
\end{eqnarray}
$\Gamma(p_1,p_2)$ is the bound state vertex function and $D(p_1,p_2)$ is given by 
\begin{eqnarray}\label{a2}
D(p_1,p_2)=D(p_1)D(p_2),
\end{eqnarray}
where $D(p)$ is the scalar field propagator, which is usually approximated by its free form as 
\begin{eqnarray}\label{a3}
D(p)=\frac{1}{p^2-m^2+i\epsilon}.
\end{eqnarray}
$I(k;p_1,p_2,\theta)$ is the interaction kernel in the non-commutative space-time. Using Weyl-Moyal correspondence, the kernel can be writen as follows 
\begin{eqnarray}\label{a4}
I(k;p_1,p_2,\theta)=\exp \left[ik\wedge (p_1-p_2)\right]I(k),
\end{eqnarray}
where $p\wedge q=\frac{1}{2}\theta^{\mu\nu}p_\mu q_\nu$
and $\theta$ is the parameter of non-commutativity \cite{witten}
\begin{eqnarray}\label{a5}
\theta^{\mu\nu}=-i[x^\mu,x^\nu].
\end{eqnarray}
In general it is not possible to find the exact solutions of the BS equation (\ref{a1}).  Therefore
we consider the ladder approximation and assume the instantaneous interactions \cite{itz}. Consequently one can rewrite the interaction kernel in the well known form
\begin{eqnarray}\label{a6}
I(k)\rightarrow I({\bf k})\sim\frac{1}{{\bf k}^2}.
\end{eqnarray}
It is shown that $\theta ^{0i}\ne 0$ lead to some problems with unitarity of field theories and the concept of causality  \cite{seiberg,gomis}.  
Therefore, we consider $\theta ^{0i}=0$. We define $E$ to be the mass of the bound state 
in the center of momentum (CM) frame. $T$ and $t$, are the CM bound state energy of the constituents. 
Thus the CM energy-momenta of the particles would be $p_1=({\bf p},t+w)$ and 
$p_2=(-{\bf p},T-w)$ and we have $E=t+T$ \cite{con}.
\par
Defining 
\begin{eqnarray}\label{d1}
\phi({\bf p})=\frac{1}{2\pi i}\int dw D(p_1,p_2)\Gamma(p_1,p_2),
\end{eqnarray}
one can show that (\ref{a1}) leads to 
\begin{eqnarray}\label{a7}
\left(({\bf p}^2+m^2)^{\frac{1}{2}}+({\bf p}^2+M^2)^\frac{1}{2}-E\right)\phi(r)=-Z({\bf p})\phi(r-\frac{i}{2}\theta.\nabla)I(r),
\end{eqnarray} 
where 
\begin{eqnarray}\label{a8}
Z({\bf p})=\frac{({\bf p}^2+m^2)^{\frac{1}{2}}+({\bf p}^2+M^2)^\frac{1}{2}}{2({\bf p}^2+m^2)^{\frac{1}{2}}({\bf p}^2+M^2)^\frac{1}{2}\left(({\bf p}^2+m^2)^{\frac{1}{2}}+({\bf p}^2+M^2)^\frac{1}{2}+E\right)}.
\end{eqnarray} 
To the first order of $\theta$, in the non-relativistic limit, (\ref{a7}) results in the 
familiar Schrodinger equation of motion for a point particle in an electromagnetic field: 
\begin{eqnarray}\label{a9}
\left[\frac{\left({\bf p}-e{\bf A}\right)^2}{2\mu}+I(r)\right]\phi=E_0\phi,
\end{eqnarray}
where 
\begin{eqnarray}\label{c1}
I(r)&=&-\frac{\alpha}{r},\nonumber\\
\label{c2}
{\bf A}&=&\frac{\mu}{2e}\theta.\nabla I(r).
\end{eqnarray}
It should be noted that ${\bf A}$ satisfies Coulomb gauge fixing condition $\nabla.{\bf A}=0$ due to antisymmetry of $\theta$. 
\par
If one defines 
\begin{eqnarray}\label{d2}
\Theta=(\theta_{23},\theta_{31},\theta_{12})
\end{eqnarray}
and
\begin{eqnarray}\label{d3}
{\bf m}=\frac{\mu}{2e}\Theta,
\end{eqnarray}
then ${\bf A}$ can be rewritten as
\begin{eqnarray}\label{b2}
{\bf A}={\bf m}\times\frac{{\bf r}}{r^3}.
\end{eqnarray}
This is similar to the vector potential field due to magnetic moment ${\bf m}$ in 
the Coulomb gauge.
Up to the order $\alpha^4$ the $\theta$-dependent term in the Hamiltonian (\ref{a9}) 
leads to an energy shift 
\begin{eqnarray}\label{a11} 
\triangle E=\left<\alpha\frac{\Theta.{\bf L}}{r^3}\right>\sim\left|\Theta\right|\alpha^4.
\end{eqnarray}
The above correction has the familiar form of the {\it normal Zeeman effect}. This result, apart from a factor $\frac{1}{4}$ has been already derived from non-commutative quantum mechanics in ref.\cite{chai}. 
Such an energy shift can impose an 
upper bound on the value of $\theta$. For instance, $\theta\sim 10^{-7}{A^{o}}^2$ 
for $\triangle E\sim 10^{-6}ev$, but to have an accurate prediction, one needs a precise 
experimental data. The hyperfine splitting of positronium (HFS) has been measured with great 
accuracy. It is known that theoretical 
predictions for HFS at the order $\alpha^6$ does not match experimental data. 
This may lead to an accurate test on the non-commutativity of space. 
Solving the BS equation for positronium in a similar way, one can easily show that 
up to the order $\alpha^4$, no spin 
dependent corrections, owing to the spatial non-commutativity, 
appeare in the positronium spectra. Therefore,  
one should calculate the higher order corrections.  
The work in this direction, using NRQED method, is in progress.
\par 
\vspace{5mm}
\par
{\bf Acknowledgement}. Financial support of Isfahan University of Technology is gratefully acknowledged.

\end{document}